\documentclass[pra,superscriptaddress,twocolumn]{revtex4-2}

\usepackage{graphicx}
\usepackage{color}
\usepackage{amsmath}
\def\vector#1{\mbox{\boldmath $#1$}}

\begin{document}
\title{Nuclear Surface Acoustic Resonance with Spin-Rotation Coupling}

\author{Koji Usami}
\email{usami@qc.rcast.u-tokyo.ac.jp}
\affiliation{Research Center for Advanced Science and Technology (RCAST), The University of Tokyo, Meguro-ku, Tokyo 153-8904, Japan}
\author{Kazuyuki Takeda}
\email{takezo@kuchem.kyoto-u.ac.jp}
\affiliation{Division of Chemistry, Graduate School of Science, Kyoto University, 606-8502 Kyoto, Japan}
\date{\today}

\begin{abstract}
We show that, under an appropriate out-of-plane static magnetic field, nuclear spins in a thin specimen on a surface acoustic wave (SAW) cavity can be resonantly excited and detected through spin-rotation coupling. Since such a SAW cavity can have the quality factor as high as $10^{4}$ and the mode volume as small as $10^{-2}$~mm$^{3}$ the signal-to-noise ratio in detecting the resonance is estimated to be quite high. We argue that detecting nuclear spin resonance of a single flake of an atomically-thin layer of two-dimensional semiconductor, which has so far been beyond hope with the conventional inductive method, can be a realistic target with the proposed scheme.
\end{abstract}

\maketitle

\textit{Introduction}.---A particle with an orbital angular momentum $\vector{L}$ in an inertial frame of reference acquires an extra energy $-\vector{L}\cdot\vector{\omega}$ in a non-inertial frame rotating with an angular velocity $\vector{\omega}$ with respect to the inertial frame~\cite{LL1}.
The same argument holds for a particle with a spin angular momentum $\vector{S}$.
Due to the extra energy $-\vector{S} \cdot \vector{\omega}$ emerged in the rotating frame as a consequence of the \textit{spin-rotation coupling}~\cite{HN1990}, the spin system is magnetized as if it were exposed to a magnetic field $\vector{B}_{\omega} = \vector{\omega} / \gamma$, where $\gamma$ is the gyromagnetic ratio of the particle.
Development of the magnetization by rotation was first observed by Barnett in 1915 in a rotating ferromagnetic body~\cite{Barnett1915}.
Very recently, the Barnett effect was also reported for paramagnetic electron spins~\cite{Ono2015} as well as for nuclear spins, causing frequency shift of nuclear magnetic resonance (NMR)~\cite{Chudo2014} and extra nuclear polarization~\cite{Arabgol2019} by sample spinning at $\sim 10$~kHz.

In this Letter, we explore the possibility of accessing nuclear spin resonance through the \textit{alternating} Barnett field in the presence of a static, polarizing magnetic field $\vector{B}_{0}$.
To this end, the Barnett field $\vector{B}_{\omega}$ has to be normal to $\vector{B}_{0}$ and be rotating around $\vector{B}_{0}$ at the frequency matched to the Larmor spin-precession frequency $\omega_{0}=- \gamma B_{0}$, which can be several tens of MHz or even higher.
To realize such seemingly impossible, rapid change of the direction of mechanical rotation and thereby of the Barnett field, we propose to exploit a surface acoustic wave (SAW) device and attach on it a thin layer of the material containing the nuclear spins of interest.
The elastic medium carrying the surface wave undergoes elliptic backward rotation~\cite{TB}, and the resultant acoustic vortex field and thereby the Barnett field oscillates at the SAW frequency. The oscillating Barnett field is a superposition of the resonantly rotating and the counter-rotating components, and the former can cause transition between the spin states, creating spin coherence that leads to a detectable back action onto the SAW device.

The presence of the spin-rotation coupling between electron spins and SAW has been predicted~\cite{Chudnovsky2007,Matsuo2013} and confirmed through generation of alternating electron-spin currents~\cite{Kobayashi2017} and through resonant excitation of spin wave~\cite{Kurimune2020} in a thin layer of conductors deposited on the SAW device.
Importantly, it is \textit{not} the magnetic moment $\gamma \vector{S}$ of the spin but its angular momentum $\vector{S}$ that is involved in the spin-rotation coupling. It follows that, for a given angular velocity $\vector{\omega}$ of mechanical rotation, the spin-rotation coupling is independent of the gyromagnetic ratio.
Therefore, even though the gyromagnetic ratios of nuclei are orders of magnitude smaller than that of electrons, the spin-rotation coupling for nuclei is expected to be comparable to that for electrons.

The proposed approach offers a new mechanism of nuclear surface acoustic resonance (NSAR), distinguishing itself from well known nuclear acoustic resonance (NAR) in bulk materials~\cite{Sundfors1983}, where the nuclear spins interact with acoustic waves through the dynamic electrical quadrupole coupling~\cite{Levitt} or dynamic Alpher-Rubin coupling~\cite{Alpher1954}.
%The former is caused by modulation of inter-atomic distances and thereby of electric-field gradient by the acoustic wave, and is relevant for quadrupolar nuclei with the spin quantum number larger than $1/2$. The latter can take place for spin-$1/2$ nuclei as well but only in metals, where the conduction electrons reacts upon the acoustic wave causing the current and thus the magnetic field that drive spin resonance.
The Barnett field induced by the SAW cavity can be confined in a volume far smaller than the size of the coil used in the conventional NMR experiments~\cite{Schuetz2015}.
Moreover, the quality factor of the state-of-art SAW cavities can reach $10^{4}$~\cite{Schuetz2015}, being two orders of magnitude higher than that of the conventional LC resonator.
The small cavity volume and the large quality factor of the SAW cavity potentially leads to the improved signal-to-noise ratio (SNR), and thereby offering a vital tool to characterize structures and dynamics of thin samples, such as van der Waals materials.

\begin{figure}[h]
\includegraphics[width=0.85\linewidth]{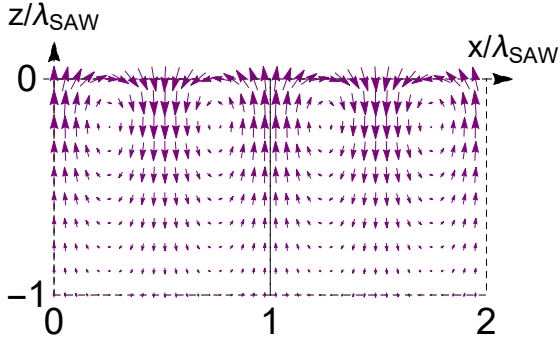}
\caption{A snapshot of a vectorial velocity field $\mathrm{Re} \left[ \dot{\vector{u}} \right]$ in the $zx$-plane accompanying with a plane monochromatic surface wave propagating along the $x$ axis. The values $q_{L}$ and $q_{T}$ used here are for LiNbO$_{3}$ with $\lambda_{\mathrm{SAW}}=40~\mathrm{\mu m}$.}
\label{fig:velocity}
\end{figure}

\textit{SAW, spin-rotation coupling, and Barnett field}.---Let us consider a semi-infinite elastic medium on which a SAW with a wavelength $\lambda_{\mathrm{SAW}}$ and an angular frequency $\omega_{0}$ propagates along the $x$ axis. The surface plane is taken to be lying in the $xy$-plane at $z=0$, and the elastic medium occupies the volume $z<0$, whereas the region $z>0$ is vacuum.
Within the monochromatic and plane-wave approximation, the displacement field $\vector{u}(x,z,t)$ is given by a sum of the longitudinal component $\vector{u}_{L} = \vector{\nabla} \psi_{0}e^{q_{L}z}e^{i \left(kx-\omega_{0} t\right)}$
and the transverse component $\vector{u}_{T} = \vector{\nabla} \times A_{0} e^{q_{T}z}e^{i \left(kx-\omega_{0} t\right)} \vector{e}_{y}$ as~\cite{TB}
\begin{eqnarray}
u_{x}(x,z,t) &=& \left( ik \psi_{0}e^{q_{L}z} - q_{T} A_{0} e^{q_{T}z} \right) e^{i\left( kx-\omega_{0}t \right)}, \label{eq:ux} \\
u_{z}(x,z,t) &=& \left( q_{L} \psi_{0} e^{q_{L}z} + ik A_{0} e^{q_{T}z} \right) e^{i\left( kx-\omega_{0}t \right)}. \label{eq:uz}
\end{eqnarray}
Here, the wave vector $k=2 \pi / \lambda_{\mathrm{SAW}}$ along the direction of propagation ($x$) is real, while those along $z$ are imaginary both for the longitudinal and the transverse displacements.
$\psi_{0}$ and $A_{0}$ are constants having units of meter$^{2}$, and depend on each other through $A_{0}=\frac{2ikq_{L}}{k^{2}+q_{T}^{2}} \psi_{0}$. Figure~\ref{fig:velocity} depicts the real part of the velocity field $\dot{\vector{u}}=\frac{\partial \vector{u}}{\partial t}$ in the $zx$-plane, where a point particle in the field undergoes elliptic backward rotation.

\begin{figure}[h]
\includegraphics[width=0.9\linewidth]{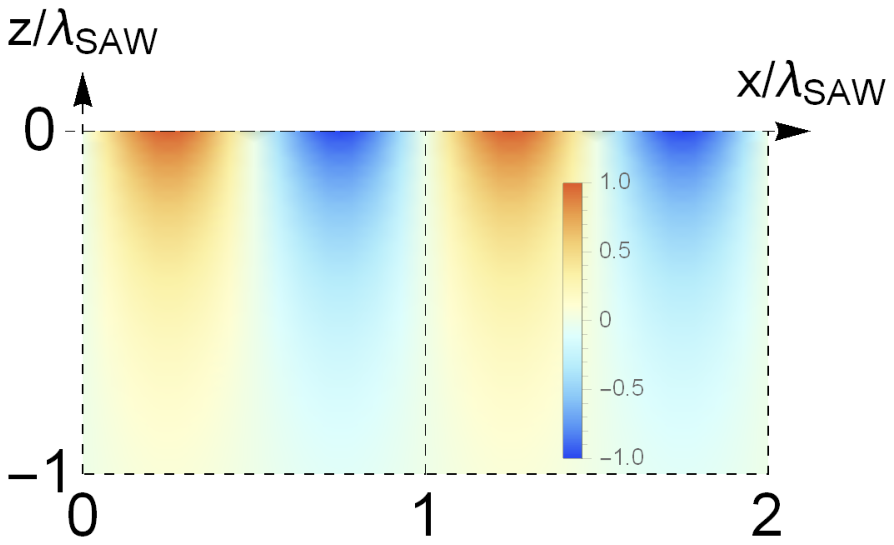}
\caption{A density plot of the $y$ component of the normalized vortex field $(\mathrm{Re} \left[ \vector{\Omega} \right])_{y} = ( \mathrm{Re} \left[\vector{\nabla} \times \dot{\vector{u}} \right])_{y}$ accompanying with the same SAW as that shown in Fig.~\ref{fig:velocity}.}
\label{fig:vortex}
\end{figure}

The vortex field $\vector{\Omega}$ accompanying the SAW is given by $\vector{\nabla} \times \dot{\vector{u}}$.
Straightforward calculation gives its dominant $y$ component $\Omega_{y}(t)$ as
\begin{equation}
  \Omega_{y}(t) =
    2 k q_{L} \omega_{0}
    \frac{k^{2} - q_{T}^{2}}{k^{2} + q_{T}^{2}} \psi_{0}
    e^{q_{T}z} e^{i(kx - \omega_{0} t)}.
\end{equation}
Figure~\ref{fig:vortex} shows a snapshot of the real part of the oscillating vortex field $\Omega_{y}(t)$, where we can observe that the field is localized in the vicinity of the surface with its amplitude decaying exponentially with $z$.
Over the lateral dimension, the amplitude is uniform, whereas the phase is alternating with $x$.
Since the angular velocity $\vector{\omega}$ is given by $\vector{\Omega}/2$, the individual nuclear spins in the thin layer on the surface of the elastic medium where the SAW propagates experience the \textit{local} spin-rotation coupling $\tilde{H}_{I} = -\frac{1}{2} \tilde{\vector{I}} \cdot \vector{\Omega}$~\cite{HN1990}, where $\tilde{\vector{I}}$ is the angular momentum density of the nuclear spins.
Alternatively, using the Barnett field $\vector{B}_{\omega} = \vector{\Omega}/(2 \gamma)$ and the nuclear magnetization $\vector{m} = \gamma \tilde{\vector{I}}$ (the magnetic moment per unit volume), the spin-rotation coupling is expressed in the form of the Zeeman coupling as
\begin{align}
\tilde{H}_{I} = -\vector{m} \cdot \vector{B}_{\omega}, \label{eq:Hi}
\end{align}
The Barnett field associated with the SAW is oscillating at the angular frequency $\omega_{0}$ of the SAW, which can be far higher than those possible with pneumatic spinning of a sample container. As a consequence, under the out-of-plane static magnetic field $\vector{B}_{0}$ nuclear spins experience resonance when $\omega_{0}= - \gamma \vector{B}_{0}$.

Since the phase of the Barnett field changes with $x$, the transverse magnetization has to change its phase with $x$ in the same way, in order to be detected by the SAW device.
This requirement is fulfilled either by employing the same SAW mode both for excitation and detection, or by creating the helical transverse magnetization using the conventional radio-frequency excitation in combination with a pulsed field gradient.

\textit{Signal-to-noise ratio}.---The SNR for the proposed NSAR detection can be analyzed by following a general formalism developed by Sidles and Rugar on the SNR of any detectors comprised of a harmonic oscillator coupled to the precessing magnetic moment~\cite{SR1993}.
Indeed, the SNR for both the conventional electrical detection and mechanical detection of NMR has successfully been described with this theoretical framework.
In the present case, the equation of motion for the Barnett field $\vector{B}_{\omega}(t) = B_{\omega}(t) \vector{e}_{y}$ in the sample on the SAW device is
\begin{equation}
m \ddot{B}_{\omega}(t) + m \frac{\omega_{0}}{Q}\dot{B}_{\omega}(t) + m \omega_{0}^{2}B_{\omega}(t) = f(t) - \vector{e}_{y} \cdot \vector{M}(t), \label{eq:eom}
\end{equation}
where $\omega_{0}$, $Q$, and $f(t)$ are the resonance angular frequency, the quality factor, and the Langevin noise of the magnetic oscillator, respectively. Here, $m$ is the \textit{magnetic mass} having units of kilogram $\times$ (meter$^{2}$/tesla$^{2}$), with which the \textit{magnetic spring constant} $\kappa_{m}$ is expressed as $\kappa_{m} = m \omega_{0}^{2}$~\cite{SR1993}.
From the equipartition theorem, the spectral density $S_{\mathrm{ff}}$ of the magnetic Langevin noise $f(t)$ is given by
\begin{equation}
S_{\mathrm{ff}}
= 4 \left( \frac{1}{\omega_{0}Q} \right) \kappa_{m} k_{\mathrm{B}}T, \label{eq:Sff}
\end{equation}
where $k_{\mathrm{B}}$ is the Boltzmann constant and $T$ is the temperature of the magnetic oscillator.
With the oscillating transverse magnetic moment, $M_{y}(t)=M_{0} \vector{e}_{y}\cos \omega_{0}t$, we have the following rms SNR~\cite{SR1993}:
\begin{equation}
\Psi_{\mathrm{rms}} = \frac{\frac{M_{0}}{\sqrt{2}}}{\sqrt{S_{\mathrm{ff}} \Delta \nu}} = M_{0} \sqrt{\frac{\omega_{0}Q}{8\kappa_{m}k_{\mathrm{B}}T \Delta \nu}}, \label{eq:SNR}
\end{equation}
where $\Delta \nu$ is the measurement band width.

To evaluate the SNR, all we need is to deduce the magnetic spring constant $\kappa_{m}$ in Eq.~(\ref{eq:SNR}). To this end, let us consider the Hamiltonian describing the SAW oscillator with a mass $\mu$, displacement $U(t)$, and momentum $P(t)$ as
\begin{equation}
H_{\mathrm{SAW}} = \frac{1}{2\mu} P(t)^{2} + \frac{1}{2} \mu \omega_{0}^{2} U(t)^{2} + H_{I}. \label{eq:Hsaw}
\end{equation}
Here, $P(t)$ is obtained by equating the kinetic energy of the oscillator to the integral of the kinetic energy density of the SAW over the volume containing a two-dimensional (2D) Gaussian SAW mode~(see Appendix~\ref{sec:Gaussian}), i.e., $\frac{1}{2 \mu} P(t)^{2} = \int_{\mathrm{cavity}} \rho \dot{\vector{u}} \dot{\vector{u}}^{*} dv$, with $\rho$ being the mass density of the elastic medium.
The last term $H_{I}$ provides the nuclear-spin--oscillator coupling, which can, from Eq.~(\ref{eq:Hi}), be read as~(see Appendix~\ref{sec:noc})
\begin{eqnarray}
H_{I} &=& \int_{\mathrm{cavity}} \tilde{H}_{I} dv = - \frac{1}{2\gamma}\int_{\mathrm{cavity}} \vector{m} \cdot \vector{\Omega} dv \nonumber \\
&=& - \frac{k}{2\gamma} \left( \frac{P(t)}{ \zeta \mu} \right) \vector{e}_{y} \cdot \vector{M}(t),
\end{eqnarray}
where $\vector{M}(t)$ is the nuclear magnetic moment within the specimen put on the SAW cavity. The cavity effectively removes the spatial degree of freedom of the SAW and makes it possible to model the SAW mode as a whole as if it were a single oscillator~\cite{GR1995}. Here, the dimensionless constant $\zeta$ is a geometrical factor ranging from $0.1$ to $10$~(see Appendix~\ref{sec:zeta}), and $\zeta^{2}$ can be interpreted as an effective mass coefficient of the oscillator~\cite{GR1995}, which, in the current context, is determined by the overlap between the SAW cavity mode and the shape of the sample~(see Appendix~\ref{sec:noc}).
The thickness of the sample on the SAW is assumed to be comparable or shorter than $\lambda_{\mathrm{SAW}}$, so that the amplitude of the Barnett field is uniform over the sample of interest.
We thus have the oscillating Barnett field associated with the SAW in the cavity as $\vector{B}_{\omega}(t) = B_{\omega}(t) \vector{e}_{y} = \left( \frac{ k P(t) }{2 \gamma \zeta \mu} \right) \vector{e}_{y}$.

To make the connection to the standard equation of motion, Eq.~(\ref{eq:eom}), we need to perform a unitary transformation~\cite{SR1993} to change the canonical variables $\{ U(t), P(t) \}$ into $\{ B_{\omega}(t), \Pi(t) \}$, namely, $B_{\omega}(t) = \frac{k}{2 \gamma \zeta \mu} P(t)$ and $\Pi(t) =  -\frac{2 \gamma \zeta \mu}{k} U(t)$.
Consequently, the Hamiltonian $H_{\mathrm{SAW}}$ in Eq.~(\ref{eq:Hsaw}) becomes
\begin{widetext}
\begin{equation}
H_{\mathrm{B}} = \frac{1}{2} \zeta^{2} \mu \left( \frac{2 \gamma}{k} \right)^{2} B_{\omega}(t)^{2}
  + \frac{1}{2} \frac{\omega_{0}^{2}}{\zeta^{2} \mu \left( \frac{2 \gamma}{k} \right)^{2}} \Pi(t)^{2}
 - B_{\omega}(t) \vector{e}_{y} \cdot \vector{M}(t),
\end{equation}
\end{widetext}
and the equation of motion for $B_{\omega}(t)$ becomes the standard form given by Eq.~(\ref{eq:eom}). Here, the magnetic spring constant $\kappa_{m}$, given by
\begin{equation}
\kappa_{m} =  \zeta^{2} \mu \left(
            \frac{2 \gamma}{k}
          \right)^{2}, \label{eq:km_B}
\end{equation}
is proportional to the gyromagnetic ratio $\gamma$ squared. From Eq.~(\ref{eq:SNR}), $\Psi_{\mathrm{rms}}$ for the proposed NSAR detection scheme grows as $\gamma^{3/2}$, unlike the conventional nuclear induction scheme where $\Psi_{\mathrm{rms}} \propto \gamma^{5/2}$. This difference comes from the fact that the former NSAR scheme addresses the angular momentum of nuclei, while the latter conventional NMR addresses the magnetic moment of nuclei.

\begin{figure}[h]
\includegraphics[width=0.9\linewidth]{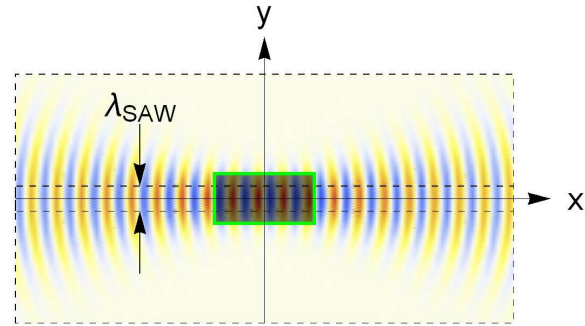}
\caption{A mode profile of a displacement field $\mathrm{Re} \left[ u_{x} \right]$ at $z=0$ and $t=0$ within a 2D Gaussian-SAW cavity mode made of LiNbO$_{3}$ with the beam waist $w_{0}=\lambda_{\mathrm{SAW}}$. Here, the area of the SAW cavity can be considered as $A = 2\lambda_{\mathrm{SAW}} \times 40\lambda_{\mathrm{SAW}}$ with $\lambda_{\mathrm{SAW}} = 40~\mathrm{\mu m}$, thus only the central region of it is depicted. Superimposed is a sample with its lateral dimension of $2 \lambda_{\mathrm{SAW}} \times 4 \lambda_{\mathrm{SAW}}$ (marked by the green rectangle).}
\label{fig:sample}
\end{figure}

The NSAR can be detected either electronically, or optically, through the SAW-cavity response. As for the electronic signal transduction an inter-digitated capacitor converts the acoustic signal into an electric signal, which is in turn amplified electronically. The optical transduction can be carried out through the acousto-optic effect such as the moving boundary effect, the photoelastic effect, and electro-optic effect~\cite{Loncar2019}. Note here that as for the latter optical scheme, the signal carried by SAW can, in principle, be faithfully transferred into an optical signal with the minimum quantum noise added~\cite{Clerk2010}. The similar idea of rf signal-to-optical signal transduction schemes based on electro-mechanical system for NMR detection have been pursued~\cite{Polzik2014,Takeda2018,Tominaga2019,Simonsen2019_1,Simonsen2019_2}.

As concrete examples, let us examine three different spin-$\frac{1}{2}$ nuclei, $^{1}$H, $^{13}$C, and $^{31}$P. As for the SAW material, we take lithium niobate (LiNbO$_{3}$) with the mass density $\rho$ of $4.65$~g/cm$^{3}$. Targeting the SAW wavelength $\lambda_{\mathrm{SAW}}$ of $\sim$~40~$\mathrm{\mu}$m, we have the resonance frequency $\omega_{0}/2 \pi$ of 88 MHz, which is determined by the dispersion relation~\cite{TB}.
Since the gyromagnetic ratios are $\gamma=268 \times 10^{6}$, $67.3 \times 10^{6}$, and $108 \times 10^{6}$~rad~$\cdot$~s$^{-1}\cdot$~T$^{-1}$ for $^{1}$H, $^{13}$C, and $^{31}$P, respectively, we need $B_{0}=$2.1, 8.2, and 5.1~T to bring the nuclear-spin-resonance frequencies to $\omega_{0}$.
The volume of the SAW cavity, $V_{c}=A \lambda_{\mathrm{SAW}}$, can be made far smaller than that of the pickup coil used for the standard inductive detection of NMR.
A SAW cavity having the area $A = 2\lambda_{\mathrm{SAW}} \times 40\lambda_{\mathrm{SAW}}$ should be feasible~\cite{Schuetz2015}. Figure~\ref{fig:sample} displays a 2D Gaussian-SAW cavity mode within the $xy$-plane with the beam waist $w_{0}=\lambda_{\mathrm{SAW}}$. To deduce the geometric factor $\zeta$, we consider a sample with its lateral dimension of $2 \lambda_{\mathrm{SAW}} \times 4 \lambda_{\mathrm{SAW}}$ and the thickness of $\lambda_{\mathrm{SAW}}/100 \sim 400$~nm, which is put on the center of the SAW cavity as shown by the green rectangle in Fig.~\ref{fig:sample}. The geometric factor $\zeta$ is then found to be $\sim 0.6$~(see Appendix~\ref{sec:zeta}). With these parameters, the magnetic spring constant, Eq.~(\ref{eq:km_B}) becomes $\kappa_{m}=9.0 \times 10^{-2}$, $0.6 \times 10^{-2}$, and $1.6 \times 10^{-2}$~J~$\cdot$~T$^{-2}$ for $^{1}$H, $^{13}$C, and $^{31}$P, respectively. These values are well-compared to the one for the inductive method with a small micro-coil~\cite{SR1993}. The quality factor of the SAW cavity could reach $10^{4}$~\cite{Schuetz2015}, which is $10^{2}$ times better than those of the good electromagnetic LC resonators ($Q \sim 10^{2}$~\cite{SR1993}), we expect the SNR $\Psi_{\mathrm{rms}}$ in Eq.~(\ref{eq:SNR}) of the NSAR could be significantly increased compared with those of the conventional NMR with the inductive method.

There are several other unconventional ways to detect NMR from small number of spins. They include the resistively-measured scheme with nano-scale point contact in fractional quantum Hall regime~\cite{Yusa2005} and the magnetic resonance force microscopy (MRFM)~\cite{Rugar2007,Rugar2009}, as well as the scheme based on nitrogen-vacancy spin magnetometers~\cite{Rugar2013,Wrachtrup2013}. The first scheme is applicable only for internal nuclei and sample-specific. The latter two methods are tuned to detect spins in the limited volume and good for three-dimensional imaging. The proposed NSAR detection scheme is unique in that it is particularly suitable for 2D thin samples.

\textit{Prospects}.---The question we now ask is to what extent can a sample be thin?
By way of illustration we shall estimate the expected SNR for the particular case of $^{77}$Se contained in WSe$_{2}$, one of 2D semiconductors called transition metal dichalcogenides~(TMDs)~\cite{XYXH2014}, for which the optically-pumped dynamic nuclear polarization (DNP) technique~\cite{SEB2017} is expected to work~\cite{Xu2013,Urbaszek2014,Marie2014}.
Take a monolayer flake of WSe$_{2}$ with the lateral size of 160~$\mu$m $\times$ 80~$\mu$m, which matches within the central region of the Gaussian mode of the 88-MHz SAW cavity shown in Fig.~\ref{fig:sample}. Obtaining such a large-area single-crystalline monolayer flake is challenging but we note the promising developments~\cite{Chhowalla2013}. With the natural abundance 7.63~\% of $^{77}$Se, the flake contains ca. $2.1 \times 10^{10}$ $^{77}$Se nuclei, which is still four orders of magnitude smaller compared with the number of spins ever successfully detected by DNP-NMR using the conventional induction method, where the single-shot rms SNR was reported to be 0.6~\cite{Barrett1994}. Under the magnetic field of $B_{0}=10.8$~T, $^{77}$Se nuclei ($\gamma=51.0 \times 10^{6}$~rad~$\cdot$~s$^{-1}\cdot$~T$^{-1}$) on the SAW cavity can be resonantly excited by the SAW. From Eq.~(\ref{eq:SNR}) with $T=4$~K and $\Delta \nu =10$~kHz, we have the single-shot rms SNR, $\Psi_{\mathrm{rms}} \sim 0.3 \times 10^{-3}$. Now, assuming 150-fold improvement of nuclear spin polarization by optically-pumped DNP, the available spins increase up to roughly $3 \times 10^{9}$ and we have $\Psi_{\mathrm{rms}} \sim 0.04$, which suggests that 1000-time average would allow us to achieve unity SNR. We anticipate the NASR scheme to bring new insight that help to understand the relatively unexplored role of nuclear spins in 2D semiconductors.

\textit{Conclusion}.---The oscillating Barnett field created by a small-volume high-Q SAW cavity can be exploited to detect nuclear spin resonance through the spin-rotation coupling. The proposed scheme is particularly well-suited to investigate nuclei in 2D extended samples and the expected SNR suggests that detection of an NSAR signal from a single flake of atomically-thin 2D semiconductor is feasible once combined with the DNP technique.

\textit{Acknowledgments.}---
We are indebted to Yasunobu Nakamura, Ryusuke Hisatomi, Kotaro Taga, Rekishu Yamazaki, Shotaro Kano, Yuichi Ohnuma, Ryo Sasaki, Toshiya Ideue, and Yoshihiro Iwasa for useful discussion. We acknowledge financial support from JST ERATO (Grant Number JPMJER1601), JST CREST (Grant Number JPMJCR1873), and JSPS KAKENHI (Grant Number 19H05602).

\appendix

\begin{widetext}

\section{Two-dimensional Gaussian surface acoustic modes}
We describe here two-dimensional (2D) Gaussian surface acoustic wave (SAW) modes. Among these modes, we are interested in the focused beam-like fundamental mode, which can have a small beam radius at the beam waist of the order of tens of micrometers. A cavity that supports the fundamental mode thus has a very small mode volume, which is instrumental in realizing large spin-rotation coupling. We begin by recapitulating the basic wave equations to equip us with the notations and all that for the discussion of the 2D Gaussian SAW modes.

\subsection{Rudimentary information}
In an elastic medium with a mass density $\rho$, bulk modulus $K$, and shear modulus $\mu_{s}$, the equation for elastic waves is given by~\cite{TB}
\begin{equation}
\rho \frac{\partial^{2} \vector{u}}{\partial t^{2}} = \left( K + \frac{1}{3} \mu_{s} \right) \vector{\nabla} \left( \vector{\nabla} \cdot \vector{u} \right) + \mu_{s} \nabla^{2} \vector{u}, \label{eq:we}
\end{equation}
where $\vector{u}$ represents the displacement vector field. In terms of the longitudinal phase speed
\begin{equation}
C_{L} = \left( \frac{K + \frac{4}{3} \mu_{s}}{\rho} \right)^{1/2},
\end{equation}
and the transverse phase speed
\begin{equation}
C_{T} = \left( \frac{\mu_{s}}{\rho} \right)^{1/2},
\end{equation}
Eq.~(\ref{eq:we}) can be rewritten as
\begin{equation}
\frac{\partial^{2} \vector{u}}{\partial t^{2}} = \left( C_{L}^{2} - C_{T}^{2} \right) \vector{\nabla} \left( \vector{\nabla} \cdot \vector{u} \right) + C_{T}^{2} \nabla^{2} \vector{u}. \label{eq:we2}
\end{equation}
Note here that unlike Maxwell equations, or rather, wave equations for electromagnetic waves, the phase speeds are far slower than the speed of wave propagation, due to that the mass density $\rho$ is finite. This fact gives rise to a longitudinal acoustic wave as well as a pair of orthogonal transverse acoustic waves in the elastic medium.

In an infinitely extended medium, the longitudinal and the transverse waves behave independently with the following respective wave equations:
\begin{equation}
\frac{\partial^{2} \vector{u}_{L}}{\partial t^{2}} = C_{L}^{2} \nabla^{2} \vector{u}_{L}, \label{eq:weL}
\end{equation}
and
\begin{equation}
\frac{\partial^{2} \vector{u}_{T}}{\partial t^{2}} = C_{T}^{2} \nabla^{2} \vector{u}_{T}. \label{eq:weT}
\end{equation}

\subsection{Surface acoustic waves}

When the medium is semi-infinite, that is to say, the medium occupies only up to $z=0$ from the bottom along the $z-$direction, surface acoustic waves (SAWs) emerge. The displacement vector field is obtained by solving Eq.~(\ref{eq:we2}) with the free boundary condition at the surface $z=0$.

\begin{figure}[h]
\includegraphics[width=0.35\linewidth]{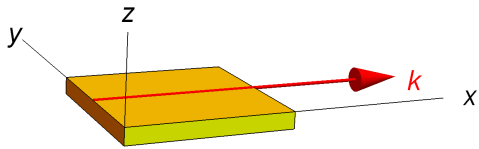}
\caption{Semi-infinite medium and a wave vector $\vector{k}$ of a SAW. }
\label{fig:geometry}
\end{figure}

\subsubsection{Plane-wave solutions}
Suppose now that a \textit{monochromatic} surface wave infinitely extended in the $y$ direction propagates in the $x$ direction, as shown in Fig.~\ref{fig:geometry}. In this case, the plane-wave solution of Eq.~(\ref{eq:we2}) is given by~\cite{TB}
\begin{equation}
\vector{u} = \left[
\begin{array}{c}
u_{x} \\
0 \\
u_{z}
\end{array}
\right] = \left[
\begin{array}{c}
i k \psi_{0} \left( e^{q_{L}z} - \frac{2 q_{L}q_{T}}{k^{2}+q_{T}^{2}} e^{q_{T}z} \right) e^{i\left(k x - \omega_{0} t \right)} \\
0 \\
q_{L} \psi_{0} \left( e^{q_{L}z} - \frac{2 k^{2}}{k^{2}+ q_{T}^{2}} e^{q_{T}z} \right) e^{i\left(k x - \omega_{0} t \right)}
\end{array}
\right], \label{eq:xi}
\end{equation}
where $\psi_{0}$ is a scalar constant, $\omega_{0}$ is the angular frequency, $k=\frac{2 \pi}{\lambda_{\mathrm{SAW}}}$ is the wave vector along $x$, and
\begin{eqnarray}
q_{L} &=& k \sqrt{1- \kappa \xi} \\
q_{T} &=& k \sqrt{1-\xi}
\end{eqnarray}
are the \textit{imaginary} wave vectors along $z$ for longitudinal and transverse modes, respectively, with
\begin{equation}
\kappa = \frac{C_{T}^{2}}{C_{L}^{2}} = \frac{1-2 \nu}{2 \left( 1-\nu \right)} \label{eq:kappa}
\end{equation}
and
\begin{equation}
\xi = \frac{C_{R}^{2}}{C_{T}^{2}} = \left( \frac{\omega}{C_{T}k}\right)^{2},
\end{equation}
and $C_{R}$ being the phase speed of the surface wave ($C_{R} < C_{T} <C_{L}$). Here, the Poisson ratio $\nu$ in Eq.~(\ref{eq:kappa}) is given in terms of $K$ and $\mu_{s}$ by
\begin{equation}
\nu = \frac{3 K -2 \mu_{s}}{2 \left( 3K + \mu_{s} \right)}.
\end{equation}

\subsubsection{Gaussian modes} \label{sec:Gaussian}
Now we consider the complications when the plane-wave condition is forgone. This can be done in two steps. The plane-wave solution for $\vector{u}$ in Eq.~(\ref{eq:xi}) can be read as a sum of the two solutions: namely, the longitudinal plane-wave solution (the plane-wave solution of Eq.~(\ref{eq:weL})), which can be obtained by taking a divergence of a \textit{scalar potential}, $\psi = \psi_{0} e^{q_{L}z} e^{i\left(k x - \omega_{0} t \right)}$, that is,
\begin{equation}
\vector{u}_{L} = \left[
\begin{array}{c}
u_{L;x} \\
0 \\
u_{L;z}
\end{array}
\right] = \vector{\nabla} \psi = \left[
\begin{array}{c}
i k \psi_{0} e^{q_{L}z} e^{i\left(k x - \omega_{0} t \right)} \\
0 \\
q_{L} \psi_{0} e^{q_{L}z} e^{i\left(k x - \omega_{0} t \right)}
\end{array}
\right], \label{eq:xiL}
\end{equation}
and the transverse plane-wave solution (the plane-wave solution of Eq.~(\ref{eq:weT})), which can be obtained by taking rotation of a \textit{vector potential}, $\vector{A}=\left[
\begin{array}{c}
0 \\
A_{0} e^{q_{T}z} e^{i\left(k x - \omega_{0} t \right)} \\
0
\end{array}
\right]$, that is,
\begin{equation}
\vector{u}_{T} = \left[
\begin{array}{c}
u_{T;x} \\
0 \\
u_{T;z}
\end{array}
\right] = \vector{\nabla} \times \vector{A}
= \left[
\begin{array}{c}
-q_{T} A_{0} e^{q_{T}z} e^{i\left(k x - \omega_{0} t \right)} \\
0 \\
ik A_{0} e^{q_{T}z} e^{i\left(k x - \omega_{0} t \right)}
\end{array}
\right], \label{eq:xiT}
\end{equation}
with the relation,
\begin{equation}
\frac{A_{0}}{\psi_{0}} = \frac{2 i k q_{L}}{k^{2}+q_{T}^{2}}, \label{eq:c}
\end{equation}
imposed from the boundary conditions.

The first step to obtain the Gaussian modes is to make the scalar constants $\psi_{0}$ and $A_{0}$ in Eqs.~(\ref{eq:xiL}) and (\ref{eq:xiT}) depending on $x$ and $y$:
\begin{eqnarray}
\psi_{0} &\rightarrow& \psi_{0}(x,y) \\
A_{0} &\rightarrow& A_{0}(x,y).
\end{eqnarray}
This step induces the changes
\begin{equation}
\vector{u}_{L} \rightarrow \vector{u}_{L} = \left[
\begin{array}{c}
\left( \frac{\partial \psi_{0}}{\partial x} + i k \psi_{0} \right) e^{q_{L}z} e^{i\left(k x - \omega t \right)} \\
\frac{\partial \psi_{0}}{\partial y} e^{q_{L}z} e^{i\left(k x - \omega t \right)} \\
q_{L} \psi_{0} e^{q_{L}z} e^{i\left(k x - \omega t \right)}
\end{array}
\right], \label{eq:xiLG}
\end{equation}
and
\begin{equation}
\vector{u}_{T} \rightarrow \vector{u}_{T} = \left[
\begin{array}{c}
-q_{T} A_{0} e^{q_{T}z} e^{i\left(k x - \omega t \right)} \\
0 \\
\left( \frac{\partial A_{0}}{\partial x} +ik A_{0} \right) e^{q_{T}z} e^{i\left(k x - \omega t \right)}
\end{array}
\right], \label{eq:xiTG}
\end{equation}
in Eqs.~(\ref{eq:xiL}) and (\ref{eq:xiT}), respectively.

The second step is to insert the \textit{ansatz}
\begin{equation}
\vector{u} = \vector{u}_{L} + \vector{u}_{T} \label{eq:a}
\end{equation}
into Eq.~(\ref{eq:we2}) assuming that the condition Eq.~(\ref{eq:c}) would still hold. Before doing this let us tidy up Eq.~(\ref{eq:we2}). Using the fact that
\begin{equation}
\frac{\partial^{2} \vector{u}}{\partial t^{2}} = -\omega_{0}^{2} \vector{u} = - C_{R}^{2} k^{2} \vector{u}
\end{equation}
and keep assuming the monochromaticity of the wave, that is $\vector{u} \propto e^{-i \omega_{0} t}$, the time dependent wave equation~(\ref{eq:we2}) becomes time-independent one
\begin{equation}
C_{R}^{2}k^{2} \vector{u} + \left( C_{L}^{2} - C_{T}^{2} \right) \vector{\nabla} \left( \vector{\nabla} \cdot \vector{u} \right) + C_{T}^{2} \nabla^{2} \vector{u} =0. \label{eq:we3}
\end{equation}
Dividing this equation by $C_{T}^{2}$ we have our version of Helmholtz's equation:
\begin{equation}
\xi k^{2} \vector{u} + \left(\frac{1}{\kappa} - 1 \right) \vector{\nabla} \left( \vector{\nabla} \cdot \vector{u} \right) + \nabla^{2} \vector{u} =0. \label{eq:we4}
\end{equation}
Inserting the \textit{ansatz} $\vector{u}$ in Eq.~(\ref{eq:a}) into Helmholtz-like Eq.~(\ref{eq:we4}), we have a partial differential equation for $\psi_{0}(x,y)$. After some lengthy manipulation, the equation can be read as
\begin{equation}
\frac{i e^{q_{L}z}}{\sqrt{1-\zeta}\sqrt{1-\kappa \zeta}} \left[ \frac{\partial^{2} \psi_{0}}{\partial y^{2}} + 2ik \frac{\partial \psi_{0}}{\partial x} \right] + \frac{2 i \kappa \ e^{q_{T}z}}{\zeta \kappa - 2}  \left[ \frac{\partial^{2} \psi_{0}}{\partial y^{2}} + 2ik \frac{\partial \psi_{0}}{\partial x} \right] = 0. \label{eq:we5}
\end{equation}
The function $\psi_{0}(x,y)$ has to satisfy this equation for any value of $z$. Thus we have
\begin{equation}
\frac{\partial^{2} \psi_{0}(x,y)}{\partial y^{2}} + 2ik \frac{\partial \psi_{0}(x,y)}{\partial x}=0. \label{eq:g}
\end{equation}
This is nothing but the familiar one having the Gaussian solutions.

\begin{figure}[h]
\includegraphics[width=0.6\linewidth]{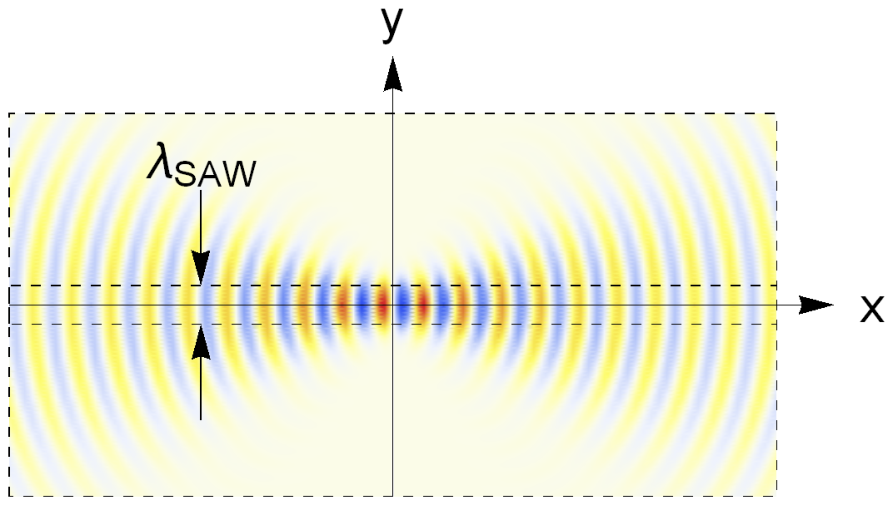}
\caption{A mode profile of a displacement field $\mathrm{Re} \left[ u_{x} \right]$ in Eq.~(\ref{eq:xix}) for the case of $w_{0}= 0.5~\lambda$ (the beam waist diameter is $\lambda$) with $\lambda_{\mathrm{SAW}} = 40~\mathrm{\mu m}$. The material is assumed to be LiNbO$_{3}$.}
\label{fig:halfL}
\end{figure}

\begin{figure}[h]
\includegraphics[width=0.6\linewidth]{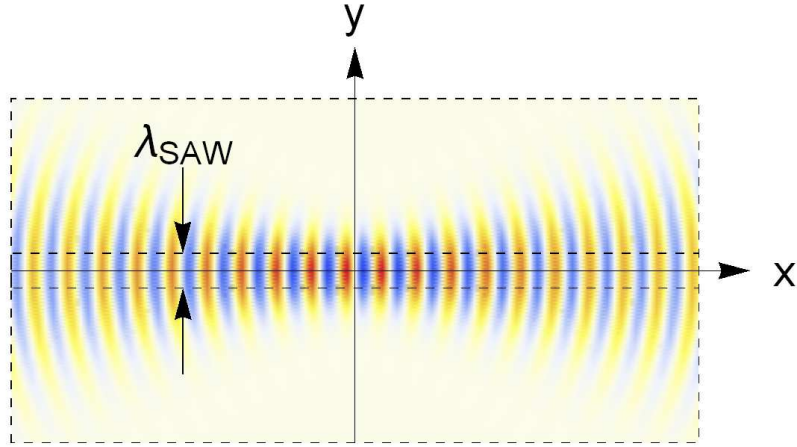}
\caption{A mode profile of a displacement field $\mathrm{Re} \left[ u_{x} \right]$ in Eq.~(\ref{eq:xix}) for the case of $w_{0}= \lambda$ (the beam diameter is 2~$\lambda$) with $\lambda_{\mathrm{SAW}} = 40~\mathrm{\mu m}$. The material is assumed to be LiNbO$_{3}$.}
\label{fig:oneL}
\end{figure}

\subsubsection{Fundamental mode} \label{sec:mode}
The solutions of Eq.~(\ref{eq:we5}) form the Gaussian modes~\cite{KL1966}. Note here that since $\psi_{0}(x,y)$ is two-dimensional it makes the form a bit different from the well-known three-dimensional Gaussian modes of electromagnetic waves. Let us now analyze the fundamental mode. Assume that the solution is given by the following Gaussian form~\cite{KL1966}:
\begin{equation}
\psi_{0}(x,y) = \alpha \exp \left[ i \left( p(x) + \frac{k}{2 q(x)} y^{2} \right) \right], \label{eq:psi0}
\end{equation}
with two \textit{complex} functions of $x$, $p(x)$ and $q(x)$, which are called the complex phase shift and the complex beam parameter, respectively. The later parameter $q(x)$ in particular plays a decisive role for the Gaussian mode. Hereafter, the amplitude of the mode $\alpha$ is considered to be normalized ($\alpha=1$). To explicitly obtain $p(x)$ and $q(x)$ let us plug the form~(\ref{eq:psi0}) into Eq.~(\ref{eq:we5}), we have two differential equations
\begin{eqnarray}
\frac{\partial q(x)}{\partial x} &=& 1 \label{eq:q} \\
\frac{\partial p(x)}{\partial x} &=& \frac{i}{2q(x)}. \label{eq:p}
\end{eqnarray}

The solution of Eq.~(\ref{eq:q}) can be written as
\begin{equation}
q(x) = q_{0} + x = -i\left( \frac{\pi w_{0}^{2}}{\lambda_{\mathrm{SAW}}} \right) + x. \label{eq:q2}
\end{equation}
Here, $q_{0}=q(x)$ is the value of the complex beam parameter at $x=0$ (\textit{beam waist}), where the value becomes pure imaginary $q_{0}=-i\left( \frac{\pi w_{0}^{2}}{\lambda_{\mathrm{SAW}}} \right)$ with $w_{0}$ being interpreted as the beam radius at the beam waist. This mystic statement can be revealed to be reasonable when we decompose the complex parameter $q(x)$ into two real parameters the \textit{radius of curvature of the wavefront} $R(x)$ and the \textit{beam radius} $w(x)$ as
\begin{equation}
\frac{1}{q(x)} = \frac{1}{R(x)} + i \left( \frac{\lambda_{\mathrm{SAW}}}{\pi w(x)^{2}} \right), \label{eq:q3}
\end{equation}
where $R(x)=\infty$ at $x=0$. With Eqs.~(\ref{eq:q2}) and (\ref{eq:q3}), we have the following useful relations:
\begin{eqnarray}
w(x) &=& w_{0} \sqrt{1+ \left( \frac{\lambda_{\mathrm{SAW}} x}{\pi w_{0}^{2}} \right)^{2}}  \\
R(x) &=& x \left[ 1+ \left( \frac{\pi w_{0}^{2}}{\lambda_{\mathrm{SAW}} x} \right)^{2} \right].
\end{eqnarray}

With the solution for $q(x)$ given by Eq.~(\ref{eq:q2}), the solution of Eq.~(\ref{eq:p}) can be written as
\begin{equation}
p(x) = \frac{i}{2} \ln \left[ 1+ i \left( \frac{\lambda_{\mathrm{SAW}} x}{\pi w_{0}^{2}} \right) \right] =  \frac{i}{2} \ln \left[ \frac{w(x)}{w_{0}} e^{i \theta(x)} \right],
\end{equation}
where
\begin{equation}
\theta(x) = \tan^{-1} \left( \frac{\lambda_{\mathrm{SAW}} x}{\pi w_{0}^{2}} \right).
\end{equation}
Putting everything together, we have the following normalized form of the fundamental mode:
\begin{equation}
\psi_{0}(x,y) = \sqrt{\frac{w_{0}}{w(x)}} \exp \left( - \frac{y^{2}}{w(x)^{2}} \right) \exp \left[ - i \left( \frac{\theta(x)}{2} - \frac{k}{2R(x)}y^{2} \right) \right]. \label{eq:psi0_2}
\end{equation}
The real part of the $u_{x}$ at $z=0$ and $t=0$ can then be obtained by inserting this form into Eq.~(\ref{eq:a}),
\begin{equation}
\mathrm{Re} \left[ u_{x} \right] = \mathrm{Re} \left[ \left( \frac{\partial \psi_{0}(x,y)}{\partial x} + i k \psi_{0}(x,y) - q_{T} \frac{2 i k q_{L}}{k^{2}+q_{T}^{2}} \psi_{0}(x,y) \right) e^{ikx} \right], \label{eq:xix}
\end{equation}
which are plotted in Figs.~\ref{fig:halfL} and \ref{fig:oneL} for the cases of $w_{0}=0.5~\lambda_{\mathrm{SAW}}$ (the beam waist diameter is $\lambda_{\mathrm{SAW}}$) and $w_{0}=\lambda_{\mathrm{SAW}}$ (the beam waist diameter is 2~$\lambda_{\mathrm{SAW}}$), respectively.

\section{Nuclear-spin--oscillator coupling}

\subsection{From SAW field to an oscillator} \label{sec:oscillator}
Let us imagine a SAW cavity carrying the fundamental mode that we have considered in Sec.~\ref{sec:mode}. The cavity effectively removes the spatial degree of freedom of the velocity field $\dot{\vector{u}}(x,y,z;t)$ and makes it possible to model the SAW mode as if it were an oscillator with a mass $\mu= \rho V_{c}$, where $\rho$ is the mass density of the elastic medium and $V_{c}$ is the cavity volume~\cite{Schuetz2015}. For the sake of concreteness we shall henceforth assume $V_{c} = T \times W \times L = \lambda_{\mathrm{SAW}} \times 2 \lambda_{\mathrm{SAW}} \times 40 \lambda_{\mathrm{SAW}}$, where $T$, $W$, and $L$ are the thickness, the width, and the length of the SAW cavity. The Hamiltonian describing the oscillator is given by
\begin{equation}
H_{\mathrm{SAW}} = \frac{1}{2\mu} P(t)^{2} + \frac{1}{2} \mu \omega_{0}^{2} U(t)^{2}, \label{eq:Hsaw0}
\end{equation}
where $U(t)$ and $P(t)$ are the displacement and the momentum of the oscillator. Here, the cycle average of $P(t)^{2}$ is obtained by equating the kinetic energy of the oscillator to the integral of the kinetic energy density of \textit{standing-wave} SAW over the SAW cavity, that is,
\begin{eqnarray}
&\ & \frac{1}{2 \mu} \langle P(t)^{2} \rangle = \frac{1}{2} \int_{\mathrm{cavity}} \rho \langle \underbrace{\left( \dot{\vector{u}}+\dot{\vector{u}}^{*} \right)}_{2 \mathrm{Re}  [\dot{\vector{u}}]} \cdot \underbrace{\left( \dot{\vector{u}}+\dot{\vector{u}}^{*}\right)}_{2  \mathrm{Re} [\dot{\vector{u}}]} \rangle dv = \int_{\mathrm{cavity}} \rho \dot{\vector{u}} \cdot \dot{\vector{u}}^{*} dv \nonumber \\
&=& \omega_{0}^{2} \rho \int_{\mathrm{cavity}} \left\{ \left( k^{2} +q_{L}^{2} \right)\left| \psi_{0} \right|^{2} e^{2 q_{L}z} +\frac{4 k^{2} q_{L}^{2}}{k^{2} + q_{T}^{2}} \left| \psi_{0} \right|^{2} e^{2 q_{T}z} - \frac{4 k^{2} q_{L} \left( q_{L} +q_{T} \right)}{k^{2} + q_{T}^{2}} \left| \psi_{0} \right|^{2}  e^{\left( q_{L} +q_{T} \right)z} + \left| \frac{\partial \psi_{0}}{\partial y} \right|^{2} e^{2 q_{L}z} \right\} dv, \nonumber \\ \label{eq:p0}
\end{eqnarray}
where $\vector{u}$ is given by Eqs.~(\ref{eq:xiLG}), (\ref{eq:xiTG}), and (\ref{eq:a}) with $\psi_{0}(x,y)$ is given by Eq.~(\ref{eq:psi0_2}). Here, we neglect the contribution from the terms with $\frac{\partial \psi(x,y)}{\partial x}$ since $\frac{\partial \psi_{0}(x,y)}{\partial x} \sim \frac{\psi_{0}(x,y)}{40 \lambda_{\mathrm{SAW}}}$ and is far less than $k \psi(x,y)$, $q_{L} \psi(x,y)$, $q_{T} \psi(x,y)$, and $\frac{\partial \psi(x,y)}{\partial y} \sim \frac{\psi_{0}(x,y)}{2 \lambda_{\mathrm{SAW}}}$ for our cavity.
From Eq.~(\ref{eq:p0}), we can assume
\begin{equation}
P(t) = 2 \mu \omega_{0} U_{c} \cos \omega_{0}t, \label{eq:p1}
\end{equation}
and
\begin{equation}
U(t) = 2 U_{c} \sin \omega_{0}t, \label{eq:u}
\end{equation}
where $U_{c}$ has a dimension of length and is defined by
\begin{eqnarray}
&\ & U_{c} =  \frac{1}{\sqrt{V_{c}}} \left[ \int_{\mathrm{cavity}} \left\{ \left( k^{2} +q_{L}^{2} \right)\left| \psi_{0} \right|^{2} e^{2 q_{L}z} +\frac{4 k^{2} q_{L}^{2}}{k^{2} + q_{T}^{2}} \left| \psi_{0} \right|^{2} e^{2 q_{T}z} - \frac{4 k^{2} q_{L} \left( q_{L} +q_{T} \right)}{k^{2} + q_{T}^{2}} \left| \psi_{0} \right|^{2}  e^{\left( q_{L} +q_{T} \right)z} + \left| \frac{\partial \psi_{0}}{\partial y} \right|^{2} e^{2 q_{L}z} \right\} dv \right]^{\frac{1}{2}}. \nonumber \\ \label{eq:Uc}
\end{eqnarray}

\subsection{Nuclear-spin--oscillator coupling} \label{sec:noc}
We now consider how the oscillator discussed in Sec.~\ref{sec:oscillator}, emerged when the SAW field has been integrated within the Gaussian SAW cavity, interacts with the nuclear spins in the sample on the SAW cavity.
The Hamiltonian for the nuclear-spin--oscillator coupling is given by integrating the interaction energy density $\tilde{H}_{I}$ over the volume of the SAW cavity, that is,
\begin{equation}
H_{I} = \int_{\mathrm{cavity}} \tilde{H}_{I} dv = - \frac{1}{2\gamma}\int_{\mathrm{cavity}}  \vector{m} \cdot \vector{\Omega} dv, \label{eq:Hi}
\end{equation}
where $\vector{m}$ is the nuclear magnetization (the magnetic moment per unit volume) and $\vector{\Omega}$ is the vortex field (for standing-wave SAW), which can be expressed as
\begin{equation}
\vector{\Omega} = \vector{\nabla} \times \underbrace{\left( \dot{\vector{u}} + \dot{\vector{u}}^{*} \right)}_{2 \mathrm{Re} [\dot{\vector{u}}]}
= \omega_{0} \left( \frac{2 k q_{L}}{k^{2} + q_{T}^{2}} \right) \left[
\begin{array}{c}
ik \frac{\partial \psi_{0}(x,y)}{\partial y} e^{q_{T}z} \\
\left( k^{2}-q_{T}^{2} \right) \psi_{0}(x,y) e^{q_{T}z} \\
q_{T} \frac{\partial \psi_{0}(x,y)}{\partial y} e^{q_{T}z}
\end{array}
\right]e^{i \left(kx-\omega_{0}t \right)} + c.c.
\end{equation}
Note that $\psi_{0}(x,y)$ given by Eq.~(\ref{eq:psi0_2}) has the maximum at $y=0$, where $\psi_{0}(x,y) \gg \frac{\partial \psi_{0}(x,y)}{\partial y}$. We thus approximate $\frac{\partial \psi_{0}(x,y)}{\partial y} \sim 0$, so that the vortex field $\vector{\Omega}(t)$ has only the $y$-component, namely, $\vector{\Omega}(t) = \Omega(t) \vector{e}_{y}$ with
\begin{equation}
\Omega(t) = 2 \omega_{0} k q_{L} \left( \frac{k^{2}-q_{T}^{2}}{k^{2} + q_{T}^{2}} \right) e^{q_{T}z} \left( \psi_{0}(x,y) e^{i \left(kx-\omega_{0}t \right)} + c.c. \right). \label{eq:vor}
\end{equation}

We note two important points here. First, the oscillating vortex field $\vector{\Omega}(t)$ along the $y$-axis can be considered as a sum of a pair of rotating components in the opposite directions. For nuclei having the positive (negative) gyromagnetic ratio $\gamma$, only the clockwise (counter-clockwise) component is relevant. Second, the phase of $\Omega(t)$, and thereby that of the rotating component, changes linearly with $x$.
This implies that a simple $\frac{\pi}{2}$ pulse created by the conventional NMR coil would not produce such transverse magnetization that is detectable with the SAW device, because the phase of the signal contribution in one place on the SAW destructively interferes with that in another. Mathematically, naively integrating the vortex field over the cavity with a uniformly distributed transverse magnetization $\vector{m}$ would result in cancellation of the integrated nuclear-spin--oscillator coupling, that is, $H_{I}=0$.

To have the non-zero nuclear-spin--oscillator coupling, the magnetization $\vector{m}$ has to be prepared in such a way that the profile of the excited magnetization constructively yields the non-zero total integrated spin-rotation coupling. We now consider two possible excitation schemes and  then the resultant nuclear-spin--oscillator coupling used for detecting the nuclear surface acoustic resonance (NSAR).

\subsubsection{Excitation scheme I} \label{sec:I}
A simple way to develop the detectable helical transverse nuclear magnetization is to use the same \textit{local} spin-rotation coupling for both the excitation and detection processes.
Initially, the nuclear magnetization is assumed to be in thermal equilibrium in the polarizing static field $\vector{B}_{0}$.
Then, the SAW cavity is excited to switch on the nuclear-spin--oscillator coupling $\tilde{H}_{I}$ given in Eq.~(\ref{eq:Hi}), so that the individual local magnetization starts to be rotated about the axis in the $xy$ plane, whose phase is determined by that of the local vortex field.
When the SAW excitation is continued until the angle of rotation has reached $\pi/2$, the $x-$dependent local magnetization becomes
\begin{equation}
\vector{m}(0) = m_{\mathrm{B}} \frac{1}{\max \left| \Omega(0) \right|} \Omega(0) \vector{e}_{x} = \frac{m_{\mathrm{B}}}{2 \psi_{0}(0,0)} e^{q_{T}z} \left( 2\mathrm{Re} \left[ \psi_{0}(x,y) e^{i kx} \right] \right) \vector{e}_{x}, \label{eq:Mt0}
\end{equation} 
where $m_{\mathrm{B}}$ is the initial thermal magnetization under the magnetic field $\vector{B}_{0}$ along $z$-axis before exciting the cavity and 
\begin{equation}
\max \left| \Omega(0) \right| = 4 \omega_{0} k q_{L} \left( \frac{k^{2}-q_{T}^{2}}{k^{2} + q_{T}^{2}} \right) \psi_{0}(0,0)
\end{equation}
is the maximum value of $\Omega(0)$ within the sample volume $V_{s}$. The magnetization $\vector{m}(0)$ is along $x$-axis and its sign is alternating as moving along $x$-axis. After the initial excitation pulse the magnetization starts precessing about $z$-axis. When ignoring relaxation, the $y$-component of the magnetization at $t$ after applying a quasi-instantaneous $\frac{\pi}{2}$-pulse at $t=0$ is given by
\begin{equation}
m_{y}(t) = m_{0} \sin \omega_{0}t, \label{eq:My}
\end{equation}
where $m_{0}$ is defined as 
\begin{equation}
m_{0} = \frac{m_{\mathrm{B}}}{2 \psi_{0}(0,0)} e^{q_{T}z} \left( 2\mathrm{Re} \left[ \psi_{0}(x,y) e^{i kx} \right] \right). \label{eq:M0}
\end{equation}
The spatial profile of the vortex field is now engraved in the magnetization $m_{y}(t)$, which will, in the end, result in the non-zero total integrated spin-rotation coupling, $H_{I}$ in Eq.~(\ref{eq:Hi}).

\subsubsection{Excitation scheme II} \label{sec:II}
An alternative scheme uses a pulsed field-gradient $B_{z}(x)$, which linearly varies with $x$. The effect of $B_{z}(x)$, applied immediately after the $\pi/2$ pulse for an interval $\tau$, is to create a periodic phase grating of the transverse magnetization, i.e., the magnetization helix in the $xy$-plane.
The condition to attain mode matching is given by
\begin{align}
  \gamma \left( \frac{\partial B_{z}}{\partial x} \right) \lambda_{\mathrm{SAW}} \tau = 2\pi.
\end{align}
For instance, a typical MRI can produce a gradient field of the order of 1 T/m.
Then, for $^{1}$H spins with $\gamma = 268 \times 10^{6}$~rad~$\cdot$~s$^{-1}~\cdot$ T, the width of the gradient pulse may be adjusted to ca. 590~$\mu$s.

\subsubsection{Detection}
Let us suppose that the transverse magnetization $m_{y}(t)$, represented as,
\begin{equation}
  m_{y}(t) = m_{0} \cos \omega_{0}t. \label{eq:MyPS}
\end{equation}
has been prepared. This can be done by changing the phase of the excitation pulse, or, considering the quasi-instantaneous $\frac{\pi}{2}$-pulse is applied at $t=-\frac{\pi}{2 \omega_{0}}$ with the excitation scheme discussed in Sec.~\ref{sec:I}.
With Eqs.~(\ref{eq:vor}) and (\ref{eq:MyPS}) the interaction Hamiltonian, Eq.~(\ref{eq:Hi}), becomes
\begin{eqnarray}
H_{I}(t) &=& - \frac{1}{2\gamma} \int_{\mathrm{sample}} \Omega(t) m_{y}(t) dv \nonumber \\
&=& - \frac{1}{2\gamma} \int_{\mathrm{sample}} 2 \omega_{0} k q_{L} \left( \frac{k^{2}-q_{T}^{2}}{k^{2} + q_{T}^{2}} \right) e^{q_{T}z} \left( \psi_{0}(x,y) e^{i \left(kx-\omega_{0}t \right)} + c.c. \right) m_{0} \cos \omega_{0}t dv. \label{eq:Hi2}
\end{eqnarray}
Here, we shall note two things: First, since the magnetization $m_{y}(t)$ in Eq.~(\ref{eq:MyPS}) is non-zero only within the sample, the integral in Eq.~(\ref{eq:Hi2}) is accordingly running only over the sample region. Second, since the magnetization is excited in such a way that its projection onto the $y$-axis is given by $m_{y}(t) = m_{0} \cos \omega_{0}t$, we can employ the \textit{rotating-wave approximation} to pickup the in-phase component, $\left(2 \mathrm{Re} \left[ \psi_{0}(x,y) e^{i kx} \right] \cos \omega_{0}t \right)$, from $\left( \psi_{0}(x,y) e^{i \left(kx-\omega_{0}t \right)} + c.c. \right)$ in Eq.~(\ref{eq:Hi2}). As a result we have, with Eq.~(\ref{eq:M0}),
\begin{eqnarray}
H_{I} &\sim& - \frac{k}{2\gamma} \left[ 2 \omega_{0} \int_{\mathrm{sample}} q_{L} \left( \frac{k^{2}-q_{T}^{2}}{k^{2} + q_{T}^{2}} \right) e^{q_{T}z} \left(2 \mathrm{Re} \left[ \psi_{0}(x,y) e^{i kx} \right] \cos \omega_{0}t \right) m_{0} \cos \omega_{0}t dv  \right] \nonumber \\
&=& - \frac{k}{2\gamma} \left[ 2 \omega_{0} \int_{\mathrm{sample}} q_{L} \left( \frac{k^{2}-q_{T}^{2}}{k^{2} + q_{T}^{2}} \right) \frac{1}{2 \psi_{0}(0,0)} \left( e^{q_{T}z} \left(2 \mathrm{Re} \left[ \psi_{0}(x,y) e^{i kx} \right] \right)^{2} \cos \omega_{0}t \right) m_{\mathrm{B}} \cos \omega_{0}t dv  \right] \nonumber \\
&=&  - \frac{k}{2\gamma} \left[ 2 \omega_{0} U_{s} \right] \cos \omega_{0}t\ \vector{e}_{y} \cdot \vector{M}(t), \label{eq:Hi3}
\end{eqnarray}
where $U_{s}$ is defined by
\begin{equation}
U_{s}= \frac{1}{V_{s}} \int_{\mathrm{sample}} q_{L} \left( \frac{k^{2}-q_{T}^{2}}{k^{2} + q_{T}^{2}} \right) \frac{1}{2 \psi_{0}(0,0)} \left( 2 e^{q_{T}z} \mathrm{Re} \left[ \psi_{0}(x,y) e^{i kx} \right] \right)^{2} dv, \label{eq:Us}
\end{equation}
having a dimension of length, and $\vector{M}(t)$ is the \textit{uniformly} oscillating magnetic moment along $y$, which is given by
\begin{equation}
\vector{M}(t) = M_{0}\vector{e}_{y} \cos \omega_{0}t
\end{equation}
with $M_{0} = m_{\mathrm{B}}V_{s}$ being the nominal magnetic moment within the sample.

With Eq.~(\ref{eq:p1}), the spin-rotation coupling Hamiltonian, Eq.~(\ref{eq:Hi3}), can then be expressed in terms of $P(t)$ as
\begin{equation}
H_{I} = - \frac{k}{2\gamma} \left( 2 \omega_{0} U_{s} \cos \omega_{0}t \right) \frac{P(t)}{2 \mu \omega_{0} U_{c} \cos \omega_{0}t} \vector{e}_{y} \cdot \vector{M}(t) = - \frac{k}{2\gamma} \left( \frac{ P(t)}{ \zeta \mu} \right) \vector{e}_{y} \cdot \vector{M}(t). \label{eq:Hi4}
\end{equation}
Here, we have introduced a dimensionless geometric factor $\zeta$ as
\begin{equation}
\zeta = \frac{U_{c}}{U_{s}},
\end{equation}
where $U_{c}$ and $U_{s}$ both have a dimension of length and are defined by Eqs.~(\ref{eq:Uc}) and (\ref{eq:Us}), respectively.

\subsection{Geometrical factor $\zeta$} \label{sec:zeta}
The geometrical factor $\zeta$ appeared in Eq.~(\ref{eq:Hi4}) can be interpreted as the square root of the effective mass coefficient of the oscillator~\cite{GR1995}, which, in the current context, is determined by the overlap between the SAW cavity mode and the shape of the sample. Figures \ref{fig:thickness}, \ref{fig:width}, and \ref{fig:length} show the geometric factors $\zeta$ as we change the thickness ($t$), the width ($w$), and the length ($l$) of the sample from the reference sample geometry, $V_{s} = t \times w \times l = \frac{\lambda_{\mathrm{SAW}}}{100} \times 2 \lambda_{\mathrm{SAW}} \times 4 \lambda_{\mathrm{SAW}}$ with $\lambda_{\mathrm{SAW}} = 40~\mathrm{\mu m}$, respectively. Here, the sample is assumed to be placed on the center of the SAW cavity whose mode profile is depicted in Fig.~\ref{fig:oneL}.

\begin{figure}[h]
\includegraphics[width=0.5\linewidth]{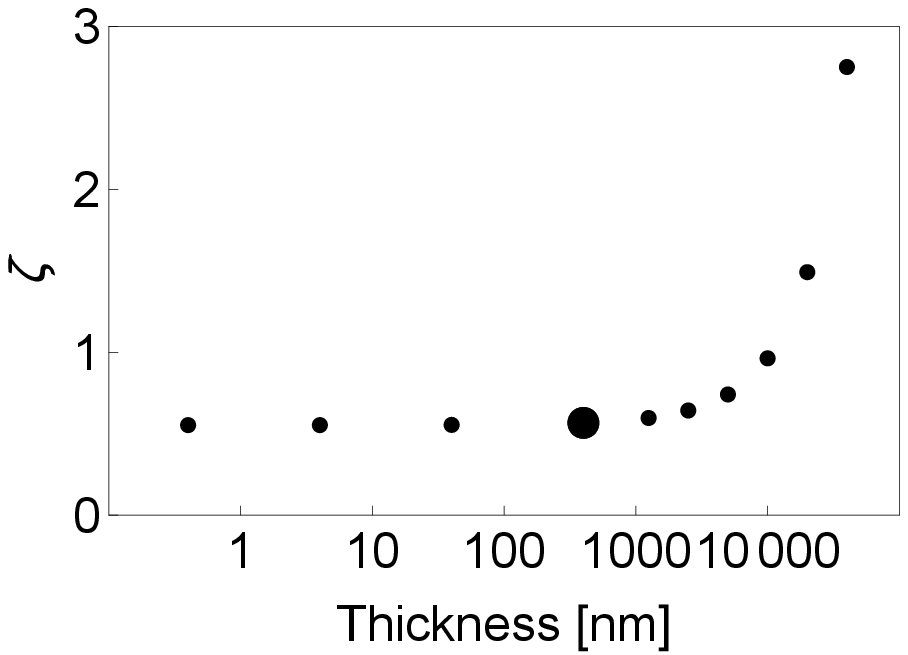}
\caption{The geometric factors $\zeta$ as a function of the thickness $t$ of the sample, where the sample geometry is $t \times 2 \lambda_{\mathrm{SAW}} \times 4 \lambda_{\mathrm{SAW}}$ with $\lambda_{\mathrm{SAW}} = 40~\mathrm{\mu m}$. Here, the sample is assumed to be placed on the center of the SAW cavity whose mode profile is depicted in Fig.~\ref{fig:oneL}. The larger point corresponds to the value $\zeta$ for  the reference sample thickness of $t = \frac{\lambda_{\mathrm{SAW}}}{100} = 400~\mathrm{n m}$.}
\label{fig:thickness}
\end{figure}

\begin{figure}[h]
\includegraphics[width=0.55\linewidth]{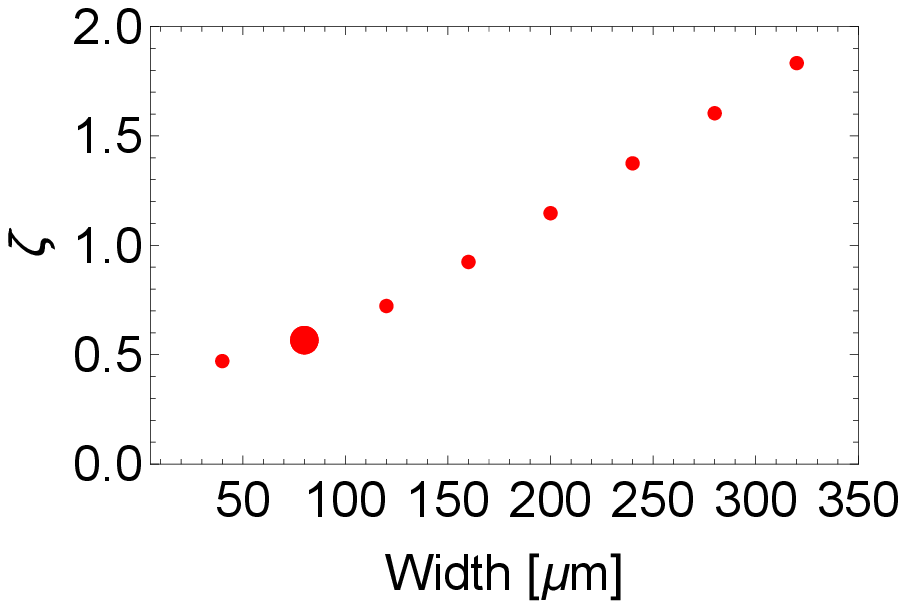}
\caption{The geometric factors $\zeta$ as a function of the width $w$ of the sample, where the sample geometry is $\frac{\lambda_{\mathrm{SAW}}}{100} \times w \times 4 \lambda_{\mathrm{SAW}}$ with $\lambda_{\mathrm{SAW}} = 40~\mathrm{\mu m}$. Here, the sample is assumed to be placed on the center of the SAW cavity whose mode profile is depicted in Fig.~\ref{fig:oneL}. The larger point corresponds to the value $\zeta$ for the reference sample thickness of $w = 2\lambda_{\mathrm{SAW}} = 80~\mathrm{\mu m}$.}
\label{fig:width}
\end{figure}

\begin{figure}[h]
\includegraphics[width=0.55\linewidth]{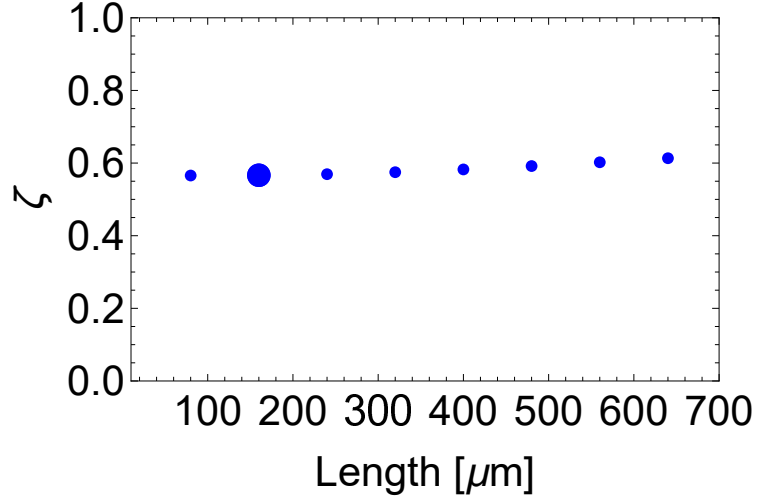}
\caption{The geometric factors $\zeta$ as a function of the length $l$ of the sample, where the sample geometry is $\frac{\lambda_{\mathrm{SAW}}}{100} \times 2 \lambda_{\mathrm{SAW}} \times l$ with $\lambda_{\mathrm{SAW}} = 40~\mathrm{\mu m}$. Here, the sample is assumed to be placed on the center of the SAW cavity whose mode profile is depicted in Fig.~\ref{fig:oneL}. The larger point corresponds to the value $\zeta$ for the reference sample thickness of $l = 4\lambda_{\mathrm{SAW}} = 160~\mathrm{\mu m}$.}
\label{fig:length}
\end{figure}

\end{widetext}


\begin{thebibliography}{99}

\bibitem{LL1}
L.~D.~Landau and E.~M.~Lifshitz, \textit{Mechanics} 3rd ed. (Butterworth-Heinemann, Oxford, 1976).

\bibitem{HN1990}
F.~W.~Hehl and W.~-T.~Ni, Phys.~Rev.~D~\textbf{42,} 2045 (1990).

\bibitem{Barnett1915}
S.~J.~Barnett,
%Magnetization by Rotation,
Phys.~Rev. \textbf{6,} 239 (1915).
%doi:10.1103/PhysRev.6.239.

\bibitem{Ono2015}
M.~Ono, H.~Chudo, K.~Harii, S.~Okayasu, M.~Matsuo, J.~Ieda, R.~Takahashi, S.~Maekawa, and E.~Saitoh,
%Barnett effect in paramagnetic states,
Phys.~Rev.~B \textbf{92,} 174424 (2015).
%doi:10.1103/PhysRevB.92.174424.

\bibitem{Chudo2014}
H.~Chudo, M.~Ono, K.~Harii, M.~Matsuo, J.~Ieda, R.~Haruki, S.~Okayasu, S.~Maekawa, H.~Yasuoka, and E.~Saitoh, Appl.~Phys.~Express~\textbf{7,} 063004 (2014).

\bibitem{Arabgol2019}
M.~Arabgol and T.~Sleator, %Observation of the Nuclear Barnett Effect,
Phys.~Rev.~Lett. \textbf{122,} 177202 (2019).

\bibitem{TB}
K.~S.~Thorne and R.~D.~Blandford, \textit{Modern Classical Physics: Optics, Fluids, Plasmas, Elasticity, Relativity, and Statistical Physics,} (Princeton University Press, Princeton, 2017).

\bibitem{Chudnovsky2007}
C.~Calero and E.~M.~Chudnovsky, Phys.~Rev.~Lett.~\textbf{99,} 047201 (2007).

\bibitem{Matsuo2013}
M.~Matsuo, J.~Ieda, K.~Harii, E.~Saitoh, and S.~Maekawa, Phys.~Rev.~B~\textbf{87,} 180402(R) (2013).

\bibitem{Kobayashi2017}
D.~Kobayashi, T.~Yoshikawa, M.~Matsuo, R.~Iguchi, S.~Maekawa, E.~Saitoh, and Y.~Nozaki,
%Spin Current Generation Using a Surface Acoustic Wave Generated via Spin-Rotation Coupling,
Phys.~Rev.~Lett.~ \textbf{119,} 077202 (2017).
%doi:10.1103/PhysRevLett.119.077202.

\bibitem{Kurimune2020}
Y.~Kurimune, M.~Matsuo, and Y.~Nozaki, Phys.~Rev.~Lett.~ \textbf{124,} 217205 (2020).

\bibitem{Sundfors1983}
R.K. Sundfors, D.I. Bolef, and P.A. Fedders, %Nuclear acoustic resonance in metals and alloys: A review,
Hyperfine Interact.~\textbf{14,} 271 (1983). %doi:10.1007/BF02043303.

\bibitem{Levitt}
M.~H.~Levitt, \textit{Spin Dynamics: Basics of Nuclear Magnetic Resonance, 2nd ed.} (John Wiley \& Sons, West Sussex, England, 2008).

\bibitem{Alpher1954}
R.A. Alpher and R.J. Rubin, %Magnetic Dispersion and Attenuation of Sound in Conducting Fluids and Solids,
J.~Acoust.~Soc.~Am.~\textbf{26,} 452 (1954).

\bibitem{Schuetz2015}
M.~J.~A.~Schuetz, E.~M.~Kessler, G.~Giedke, L.~M.~K.~Vandersypen, M.~D.~Lukin, and J.~I.~Cirac, Phy.~Rev.~X~\textbf{5,} 031031 (2015).

\bibitem{SR1993}
J.~A.~Sidles and D.~Rugar, Phys.~Rev.~Lett.~\textbf{70,} 3506 (1993).

\bibitem{GR1995}
A.~Gillespie and F.~Raab, Phys.~Rev.~D~\textbf{52,} 577 (1995).

\bibitem{Loncar2019}
L.~Shao, M.~Yu, S.~Maity, N.~Sinclair, L.~Zheng, C.~Chia, A.~Shams-Ansari, C.~Wang, M.~Zhang, K.~Lai, and M.~Lon\v{c}ar, Optica~\textbf{6,} 1498 (2019).

\bibitem{Clerk2010}
A.~A.~Clerk, M.~H.~Devoret, S.~M.~Girvin, F.~Marquardt, and R.~J.~Schoelkopf, Rev.~Mod.~Phys. \textbf{82,} 1155 (2010).

\bibitem{Polzik2014}
T.~Bagci, A.~Simonsen, S.~Schmid, L.~G.~Villanueva, E.~Zeuthen, J.~Appel, J.~M.~Taylor, A.~S\/{o}rensen, K.~Usami, A.~Schliesser, and E.~S.~Polzik, Nature~\textbf{507,} 81 (2014).

\bibitem{Takeda2018}
K.~Takeda, K.~Nagasaka, A.~Noguchi, R.~Yamazaki, Y.~Nakamura, E.~Iwase, J.~M.~Taylor, and K.~Usami, Optica~\textbf{5,} 152 (2018).

\bibitem{Tominaga2019}
Y.~Tominaga, K.~Nagasaka, K.~Usami, and K.~Takeda, J.~Magn.~Reson. \textbf{298,} 6 (2019).

\bibitem{Simonsen2019_1}
A.~Simonsen, S.~A.~Saarinen, J.~D.~Sanchez, J.~H.~Ardenkj{\ae}r-Larsen, A.~Schliesser, and E.~S.~Polzik, Opt.~Express \textbf{27,} 18561 (2019).

\bibitem{Simonsen2019_2}
A.~Simonsen, J.~D.~S\'{a}nchez-Heredia, S.~A.~Saarinen, J.~H.~Ardenkj{\ae}r-Larsen, A.~Schliesser, and E.~S.~Polzik, Sci.~Rep.~\textbf{9,} 18173 (2019).

\bibitem{Yusa2005}
G.~Yusa, K.~Muraki, K.~Takashina, K.~Hashimoto, and Y.~Hirayama, Nature~\textbf{434,} 1001 (2005).

\bibitem{Rugar2007}
H.~J.~Mamin, M.~Poggio, C.~L.~Degen, and D.~Rugar, Nat.~Nanotech.~\textbf{2,} 301 (2007).

\bibitem{Rugar2009}
C.~L.~Degen, M.~Poggio, H.~J.~Mamin, C.~T.~Rettner, and D.~Rugar, Proc.~Natl.~Acad.~Sci.~U.S.A~\textbf{106,} 1313 (2009).

\bibitem{Rugar2013}
H.~J.~Mamin, M.~Kim, M.~H.~Sherwood, C.~T.~Rettner, K.~Ohno, D.~D~Awschalom, and D.~Rugar, Science.~\textbf{339,} 557 (2013).

\bibitem{Wrachtrup2013}
T.~Staudacher, F.~Shi, P.~Pezzagna, J.~Meijer, J.~Du, C.~A.~Meriles, F.~Reinhard, and J.~Wrachtrup, Science.~\textbf{339,} 561 (2013).

\bibitem{XYXH2014}
X.~Xu, W.~Yao, D.~Xiao, and T.~F.~Heinz, Nat.~Phys.~\textbf{10,} 343 (2014).

\bibitem{SEB2017}
G.~Sharma, S.~E.~Economou, and E.~Barnes, Phys.~Rev.~B \textbf{96,} 125201 (2017).

\bibitem{Xu2013}
A.~M.~Jones, H.~Yu, N.~J.~Ghimire, S.~Wu, G.~Aivazian, J.~S.~Ross, B.~Zhao, J.~Yan, D.~G.~Mandrus, D.~Xiao, W.~Yao, and X.~Xu, Nat. Nanotechnol.~\textbf{8,} 634 (2013).

\bibitem{Urbaszek2014}
G.~Wang, L.~Bouet, D.~Lagarde, M.~Vidal, A.~Balocchi, T.~Amand, X.~Marie, and B.~Urbaszek, Phys.~Rev.~B~\textbf{90,} 075413 (2014).

\bibitem{Marie2014}
C.~R.~Zhu, K.~Zhang, M.~Glazov, B.~Urbaszek, T.~Amand, Z.~W.~Ji, B.~L.~Liu, and X.~Marie, Phys.~Rev.~B~\textbf{90,} 161302(R) (2014).

\bibitem{Chhowalla2013}
M.~Chhowalla, H.~S.~Shin, G.~Eda, L.~-J.~Li, K.~P.~Loh, and H.~Zhang, Nature Chem.~\textbf{5,} 263 (2013).

\bibitem{Barrett1994}
S.~E.~Barrett, R.~Tycko, L.~N.~Pfeiffer, and K.~W.~West, Phys.~Rev.~Lett.~\textbf{72,} 1368 (1994).

\bibitem{KL1966}
H.~Kogelnik and T.~Li, Appl.~Opt.~\textbf{5,} 1550 (1966).
　
\end{thebibliography}
\end{document}